\documentclass[aps,prb,showpacs,twocolumn,groupedaddress]{revtex4-1}

\usepackage{graphicx,graphics,color,epsfig}
\usepackage{amsmath}
\usepackage{amssymb}
\usepackage{amsfonts}
\usepackage{amsfonts}
\usepackage{epstopdf}
\usepackage{bm}
\usepackage{times,xspace}
\usepackage{color}

\begin{document}

\title{Imprints of spin-orbit density wave in the hidden order state of URu$_2$Si$_2$}

\author{Tanmoy Das}
\affiliation{Theoretical Division, Los Alamos National Laboratory, Los Alamos, New Mexico 87545 USA.}

\date{\today}

\begin{abstract}
 The mysterious second order quantum phase transition, commonly attributed to the `hidden-order' (HO) state, in heavy-fermion metal URu$_2$Si$_2$ exhibits a number of paradoxical electronic and magnetic properties which cannot be associated with any conventional order parameter. We characterize and reconcile these exotic properties of the HO state based on a spin-orbit density wave order (SODW), constructed on the basis of a realistic density-functional theory (DFT) band structure. We quantify the nature of the gapped electronic and magnetic excitation spectrum, in agreement with measurements, while the magnetic moment is calculated to be {\it zero} owing to the spin-orbit coupling induced time-reversal invariance. Furthermore, a new collective mode in the spin-1 excitation spectrum is predicted to localize at zero momentum transfer in the HO state which can be visualized, for example, by electron spin resonance (ESR) at zero magnetic field or polarized inelastic neutron scattering measurements. The results demonstrate that the concomitant broken and invariant symmetries protected SODW order not only provides insights into numerous nontrivial hidden-order phenomena, but also offers a parallel laboratory to the formation of a topologically protected quantum state beyond the quantum spin-Hall state and Weyl semimetals.
\end{abstract}
\maketitle

\section{Introduction}

The `hidden-order' (HO) state of heavy-fermion compound URu$_2$Si$_2$, which is defined by a second order phase transition, represents a new state of matter with peculiar electronic, and magnetic properties.\cite{Cv_Palstra,Mydosh_review} The particularly intriguing characteristics of the HO state are large anomaly in spin-related measurements including inelastic neutron scattering (INS),\cite{INS2,INS} nuclear magnetic resonance (NMR),\cite{NMR} magneto-torque,\cite{Matsuda} and magnetic susceptibility,\cite{susceptibility} despite the absence of any accountable static magnetic moment,\cite{zeromoment} and time-reversal symmetry breaking. These trademark signatures of the HO state $-$ which apparently conflict with each other according to the conventional physics of quantum phase transition $-$ demand a new mechanism of quantum order which concurrently involves broken and invariant symmetries. The present theoretical efforts\cite{DFT,TheoryKotliar,TheoryPepin,TheorySasha,TheoryFujimoto,TheoryHYKee,TheoryIkeda,Riseborough,TheoryHastatic}
encompassing a large body of new order parameters have been inclusive.\cite{Mydosh_review}

Essential clues to the paradoxical symmetry properties of the HO state can be lent from the theory of symmetry-protected quantum spin Hall\cite{SCZSHE} and topological phases\cite{CKane,SCZhang} of matter which have been shown in recent years to arise  from spin-orbit coupling (SOC) entanglement in the single-electron wavefunction. Another interesting analogy can be made with the theory of Weyl semimetals, originally refereed as `accidental degeneracy in bands'\cite{Weyl}, in which multiple relativistic Dirac points are scattered in the momentum space (and thus termed `semimetal') owing to a broken crystal symmetry, but they remain topologically protected via other subjective symmetry invariance.\cite{Murakami} These compelling physical concepts led us to propose and formulate an electronic interaction induced spin-orbit density wave (SODW) as an emergent phase of matter which stabilizes in SOC systems via broken translational but invariant time-reversal symmetry.\cite{DasSR,DasPRL}

The consistency of the SODW properties with many experimental signatures of the HO phase in URu$_2$Si$_2$ is appealing, particularly as it reconciles the apparently contradictory magnetic properties of this phase. Among them, we highlight the following robust properties. (1) The present order breaks translational symmetry,\cite{INS}, but thanks to SOC, the order parameter respects time-reversal symmetry and charge conservation symmetry at each lattice point. (2) As a result, no magnetic moment is induced, in agreement with measurements.\cite{zeromoment} (3) Time-reversal symmetry invariance fosters magnetic field, in addition to temperature, to be a perturbation to the order parameter, and the corresponding critical field primarily depends on the gap value and the $g$ factor. Experiments confirm the vanishing of the HO around $B\sim35$~T.\cite{magfield}. (4) As a consequence, as field increases the HO gap decreases (given other parameters such as temperature being constant), and thus the resistivity decreases. In other words, a negative magnetoresistance effect is expected within the SODW framework, when no other field-induced phase is involved. Here we explore and expand this chart to demonstrate that the SODW gives a unified explanation to the itinerant gapping of the electronic structure, and the magnetic excitation spectrum in terms of the dispersion in both energy and momentum space. Finally, we offer a detectable prediction of a second spin-1 collective mode localized inside the gap at ${\bf q}\sim 0$ in the SODW state which can be probed via electron spin resonance (ESR) or nuclear quadrupole resonance (NQR)  at zero magnetic field or polarized neutron scattering measurements.

The rest of the paper is arranged as follows. In Sec.~II, we present our first-principles bandstructure calculation for URu$_2$Si$_2$ and compute the bare susceptibility to understand the nesting condition. Here, we also present the analytical Hamiltonian of the SODW order based on two orbitals with atomistic spin-orbit coupling. We obtain the wavefunction of the SODW and show explicitly how magnetic moment vanishes for all spin channels in the SODW state. In Sec.~III, we show representative results of electronic structure and magnetic structure of the SODW in the hidden order state of URu$_2$Si$_2$. Finally we conclude in Sec.~IV.

\begin{figure*}[top]
\rotatebox[origin=c]{0}{\includegraphics[width=1.99\columnwidth]{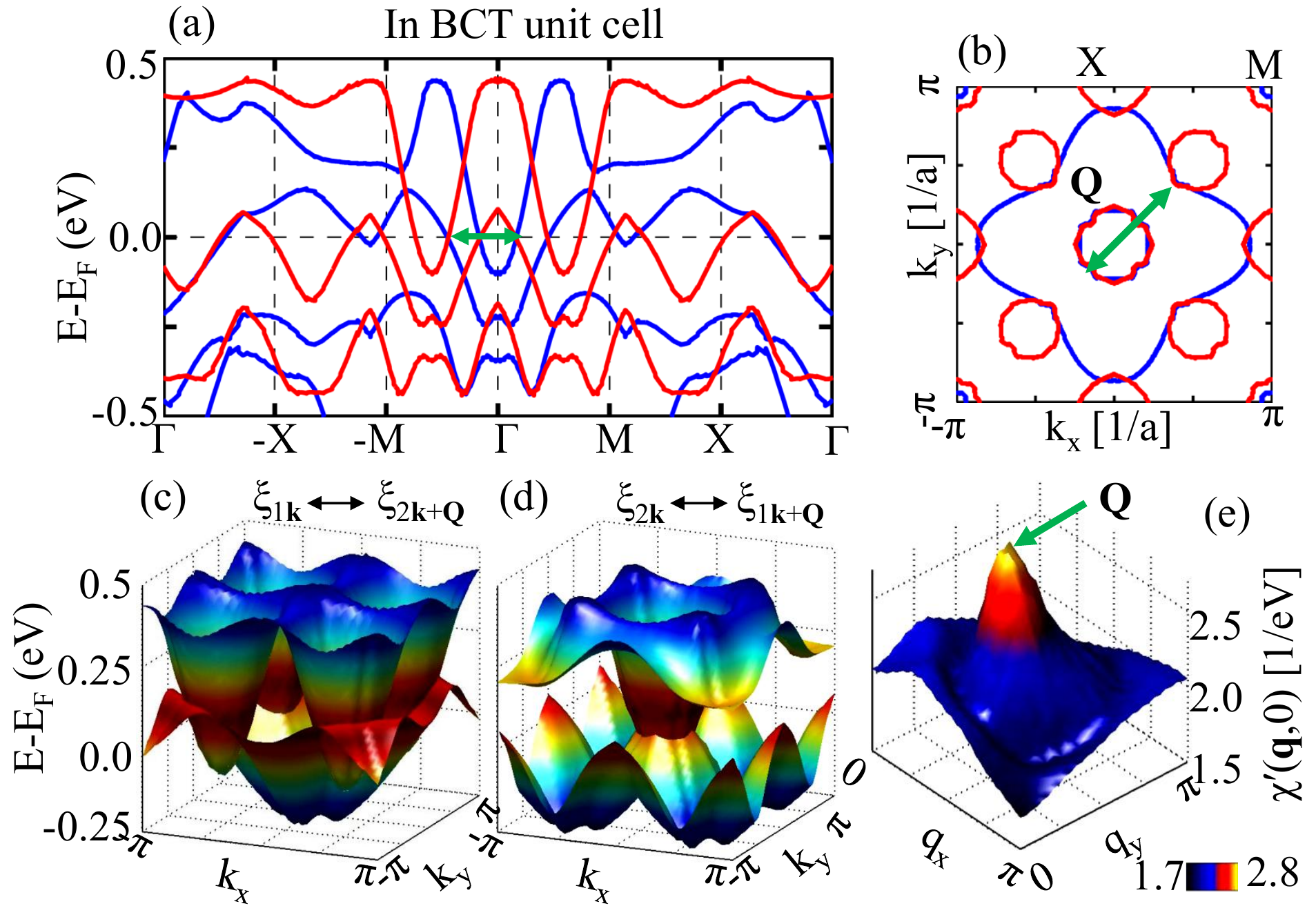}}
\caption{{FS degeneracy, and the HO `hot-spot'.} (a) Noninteracting band dispersion (blue lines) is plotted along several representative high-symmetry momentum cuts. Red lines are superimposed bands shifted by ${\bf Q}=(\pi/2, \pi/2, 0)$ in the BCT unit cell. While a significant region of nesting is visible on the FSs in (b) between original and folded bands, along the high-symmetry lines here, the dispersions exhibit linear contacts at the Fermi level with Dirac cones. This implies that the ${\bf Q}$ nesting causes the development of a density wave in the particle-hole channel with topological properties. (b) FS on $k_z$=0 plane is shown in blue line. Red line is the FS shifted by ${\bf Q}$ vectors. Good nesting condition is evident along the (110)-direction between different orbital states. This diagonal nesting reorients itself along the (100)-direction in the simple tetragonal (ST) lattice.\cite{DasSR} Some additional wiggling in the FS lines comes from the finite resolution in the calculation. (c), (d) Full non-interacting band dispersion in the entire basal plane for the two relevant nested bands gives a comprehensive view of the similarity of the FS degeneracy to the `accidental band degeneracy' for the Weyl systems studied before.\cite{Weyl,Iridates} The blue to red color map has no special significance here. (e) Static susceptibility in the 2D {\bf q} space, affirming a paramount FS instability peak at ${\bf Q}$.}
\label{fig1}
\end{figure*}

\section{Computations details.}
The low-energy electronic states of URu$_2$Si$_2$ can be defined by itinerant quasiparticle dynamics, with the inclusion of a relevant mass renormalization, because the transformation from localized to itinerant electrons occurs at a much higher temperature $\sim$60-80~K, signaled consistently by specific heat, magnetic susceptibility,\cite{Cv_Palstra} and tunneling\cite{STM,STM2} measurements. Furthermore, pertaining to the larger atomic size of 5$f$ electrons in U atoms than in the highly localized smaller 4$f$ atoms, the inter-atomic electron hopping is larger here and mobile 5$f$ electrons stem within the single-electron picture. In other words, the electronic structure of this actinide compound can be described adequately by first-principles calculation, without invoking Kondo-like physics. In fact, both DFT calculation\cite{DFT} and angle-resolved photoemission spectroscopy (ARPES) measurement\cite{Durakiewicz} have ruled out the presence of $d$ electrons within the $\pm$500~meV vicinity of the Fermi level to cause any significant hybridizations. Of course, among the octet and sextet multiplets of 5$f$ electrons, variable renormalizations to different states can cause mixing between localized and itinerant electrons within the $f$ states.

The above observation is consistent with the DFT bandstructure, calculated by including SOC using the Wien2k software,\cite{Wien,GGA} and shown by blue solid line in Fig.~\ref{fig1}(a) along several representative high-symmetry momentum ({\bf k}) directions in the body-centered tetragonal (BCT) lattice of the crystal. From their corresponding band characters, two facts emerge that the bands are dominated by 5$f$ multiplets in the low-energy scales of present interests, and these $f$ states are itinerant. The corresponding Fermi surface (FS) topology is depicted in Fig.\ref{fig1}(b), overlayed by a shifted FS with the `hot-spot' wavevector ${\bf Q}=(\pi/2,\pi/2,0)$ (red lines) in this unit cell notation. Paramount FS nesting is evident here between different orbitals, and hence identified it to be responsible for the SODW state. The FS nesting property is confirmed by static susceptibility calculation, plotted in the basal plane momentum space, in Fig.~\ref{fig1}(e), exhibiting a dominant peak at this ${\bf Q}$ vector. The band structure shifted by ${\bf Q}$ is then plotted in Fig.~\ref{fig1}(a) in red color which reveals that the low-energy bands are linear in momentum, and the contact points between main bands and folded bands at the Fermi level are in the particle-hole channel along the (110) direction. Therefore, the nesting condition in this system resembles the so-called `accidental degenerate points' or Weyl-type Dirac cones\cite{Weyl} as visualized in the entire two-dimensional momentum space in Figs.~\ref{fig1}(c) and \ref{fig1}(d) for the relevant bands. This fact justifies our ansatz that the SODW is a protected state, as in topological Weyl semimetals.

\begin{figure*}[top]
\rotatebox[origin=c]{0}{\includegraphics[width=1.9\columnwidth]{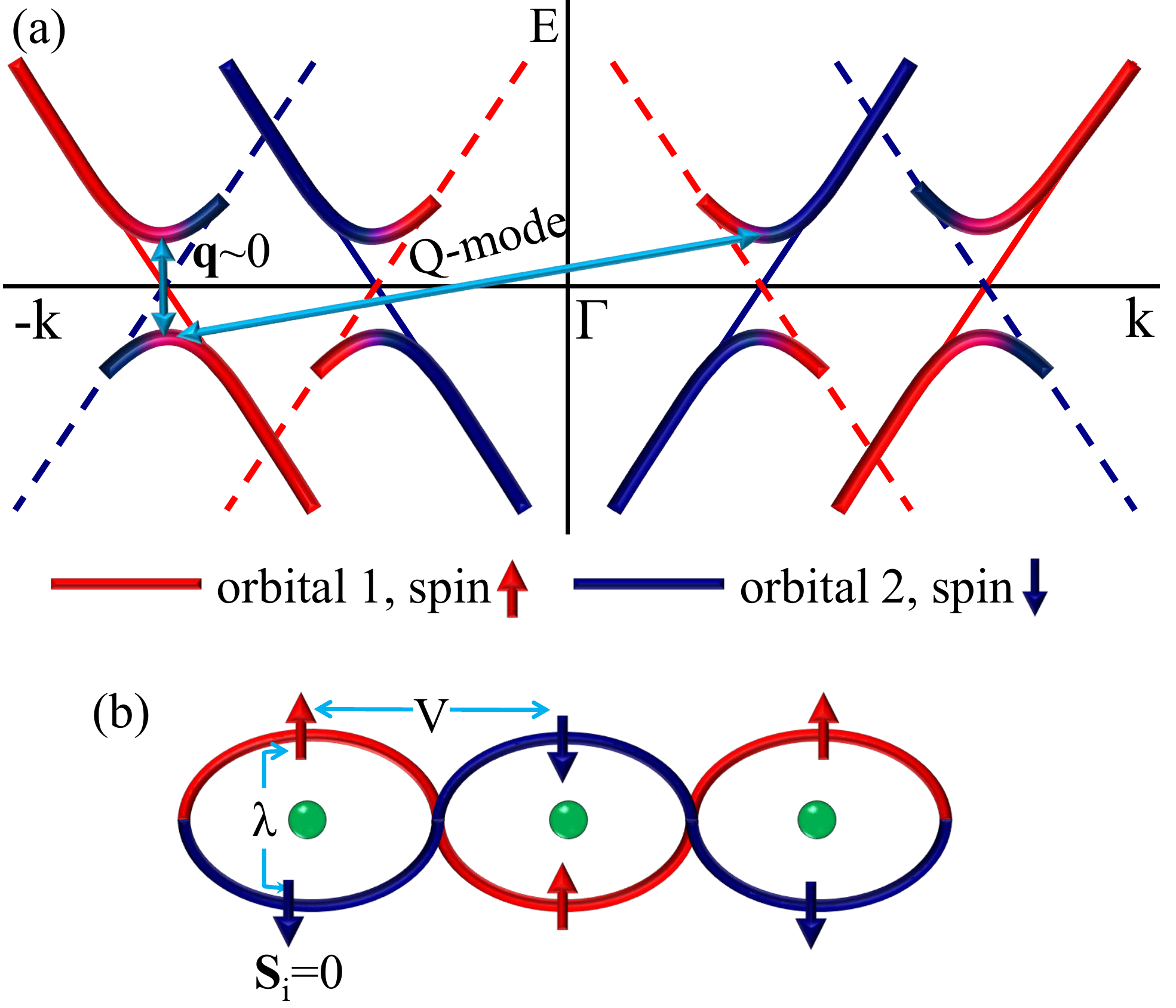}}
\caption{{Schematic illustration of SODW in momentum and real space.}  (a) The distribution of different orbital weights and corresponding entanglement with spin in the SODW state in the URu$_2$Si$_2$ electronic structure are illustrated. Due to the single-particle SOC, each orbital is spin polarized, but they mix in the SODW state determined by the coherence factors. As shown in Fig.~\ref{fig1}, two orbital states along the (110) direction in the BCT lattice are nested, and thus becomes gapped at the Fermi level. Due to the relevant density wave coherence factors, the quadratic gapped bands on both sides of the Fermi level share both spin-orbital weights, or strictly speaking different total angular momentum $J$ in the U atom, as indicated by the red to blue gradient colormap. Since the spin-flip commences in different orbitals, a direct spin-flip transition with no momentum transfer (${\bf q}\sim0$) is turned on via the umklapp scattering process in this scenario without breaking time-reversal symmetry, as computed in Fig.~\ref{fig4}. A collective mode at ${\bf Q}$ also appears between different orbitals. (b) A real-space view of the SODW. Here the nearest atomic distance is between two sublattices at which the SODW order parameter modulates due to onsite interaction $V$. But owing to atomic spin-orbit coupling in each site, counter spin polarization in the other orbital state is present (denoted by $\lambda$) which makes the total spin moment vanish at each spatial point, and the Hamiltonian preserves time-reversal invariance. Of course, perturbations such as crystal distortion or orbital fluctuations may render a finite magnetic moment at each site in this setup, but its magnitude can be expected to be small.}
\label{fig2}
\end{figure*}

Constrained by the FS properties and the nesting condition, it is appropriate to write down the effective theory for the HO state by using two nested bands. Furthermore, although we take the bandstructure directly from above the DFT calculation to obtain numerical results in Sec. III, to explicate the physical mechanism of SODW and the role of SOC in it, we start from a two-orbital basis with atomistic SOC, $\lambda$. The purpose here is to derive a general Hamiltonian of such kind and obtained analytical solution of the SODW wavefunction and nonmagnetic ground state, in which no particular assumption of the orbital symmetry and the relevant details of the SOC are imposed. Our starting noninteracting Hamiltonian is thus
\begin{equation}
H=\sum_{j{\bf k},\sigma=-{\bar\sigma}}\left[\xi_{j\bf k}c^{\dag}_{j{\bf k},\sigma}c_{j{\bf k},\sigma} + i\sigma\lambda c^{\dag}_{j{\bf k},\sigma}c_{j{\bf k},{\bar{\sigma}}}\right],
\label{H0}
\end{equation}
where $c^{\dag}_{j{\bf k}\sigma}$ is the electron creation operator in the $j^{th}$ orbital with spin $\sigma=\pm$, and $\xi_{j\bf k}$ is its corresponding dispersion spectrum in a translationally invariant crystal of Bloch momentum ${\bf k}$. With a translational symmetry breaking at the reduced reciprocal wavevector or the ``hot-spot'' vector ${\bf Q}$, the new spin-orbit basis in the Nambu-space yields $\Psi_{\bf k}= (c_{1{\bf k},\uparrow},~c_{2{\bf k},\downarrow},~c_{1{\bf k}+{\bf Q},\uparrow},~c^{\dag}_{2{\bf k}+{\bf Q},\downarrow})$. The spin-polarized Nambu operator is an important consideration which distinguishes the present SODW order parameter from an inter-orbital spin-density wave (see, for example, Ref.~\onlinecite{Riseborough}), and promotes a {\it zero} magnetic moment as will be evaluated rigorously below. In this basis, we express our interacting Hamiltonian in a matrix form:
\begin{eqnarray}
H_{\bf k}=\left(
\begin{array}{cccc}
\xi_{1{\bf k}} & i\lambda  & 0 & \Delta \\
-i\lambda & \xi_{2{\bf k}} & \Delta^* & 0\\
0        &  \Delta   & \xi_{1{\bf k}+{\bf Q}} & -i\lambda \\
\Delta^* &  0        & i\lambda & \xi_{2{\bf k}+{\bf Q}}\\
\end{array}
\right).
\label{Hint}
\end{eqnarray}
The SODW field parameter $\Delta$ is taken to be complex for generality and will be defined below. We diagonalize the above Hamiltonian in two steps by the Bogolyubov method. First, we define the Bogolyubov spinor for the SOC term as $(b_{1{\bf k},\sigma},~b_{2{\bf k},\bar{\sigma}})^{\dag}=\alpha_{\bf k}\mathbb{I}-\sigma\beta_{\bf k}i{\tau}^y (c_{1{\bf k},\sigma},~c_{2{\bf k},\bar{\sigma}})^{\dag}$, where $\mathbb{I}$ is the 2$\times$2 identity matrix, and $\tau_y$ is the second Pauli matrix, representing orbital basis. After this basis transformation, the Hamiltonian in Eq.~\ref{H0} becomes diagonal with eigenstates $E_{1,2{\bf k},\sigma}=\xi^+_{\bf k}\pm E_{0{\bf k}}$, and the corresponding spin-orbit helical weight is $\alpha^2_{\bf k} (\beta^2_{\bf k})=\frac{1}{2}\left(1\pm\frac{\xi^-_{\bf k}}{E_{0\bf k}}\right)$, where $\xi^{\pm}_{\bf k}=(\xi_{1\bf k}\pm\xi_{2\bf k})/2$, and $E^2_{0{\bf k}}=(\xi^-_{\bf k})^2+\lambda^2$. For our main computation and presentation, we start with this spin-orbit basis accessed directly from the DFT calculation including SOC and include interaction.

The subsequent diagonalization onto the Nambu-basis on the reduced Brillouin zone follows similarly, and the spin and orbital notations are retained explicitly. Here the Bogolyubov basis transformation is $(d_{n{\bf k},\sigma},~d_{m{\bf k}+{\bf Q},\bar{\sigma}})^{\dag}=u_{{\bf k}}\mathbb{I}-\nu_{nm}\sigma v_{{\bf k}}i{\tau}^y (b_{n{\bf k},\sigma},~b_{m{\bf k}+{\bf Q},\bar{\sigma}})^{\dag}$, where $n=1,2\ne m$ are the orbital indices, and $\nu_{12}=-\nu_{21}=1$ is the spin-orbit helicity index which changes sign when orbitals are interchanged. Here the canonical density operators $u_{{\bf k}} (v_{{\bf k}})$ take equivalent forms as that of $\alpha_{\bf k}/\beta_{\bf k}$ defined before, after replacing the corresponding quasiparticle states with $\tilde{\xi}^{\pm}_{\bf k}=(E_{1\bf k}\pm E_{{2\bf k}+{\bf Q}})/2$, and  $\tilde{E}^2_{0{\bf k}}=(\tilde{\xi}^-_{\bf k})^2+|\Delta|^2$. The final quasiparticle spectra in the SODW state with SOC are thus ${\tilde E}^{\pm}_{1,2{\bf k},\sigma}={\tilde \xi}^+_{\bf k}\pm {\tilde E}_{0{\bf k}}$, and two more eigenvalues when orbitals 1 and 2 are interchanged. A schematic view of the SODW split bands and the shared orbital character, determined by coherence factors $u_{\bf k}/v_{\bf k}$ between them is given in Fig.~\ref{fig2}(a). It is interesting to notice that the SOC allows the many-particle wavefunction to have a dynamical spin flip on the same orbital at the same momentum as indicated by vertical arrow, which governs a collective ${\bf S}=1$ mode to be derived below.

It is worthwhile to combine the above two steps to visualize how the four-vector SODW Nambu operator transforms under translational symmetry breaking while retaining time-reversal symmetry:
\begin{widetext}
\begin{eqnarray}
\left(\begin{array}{c}
c_{1{\bf k},\sigma}\\
c_{2{\bf k},\bar{\sigma}}\\
c_{1{\bf k}+{\bf Q},\sigma}\\
c_{2{\bf k}+{\bf Q},\bar{\sigma}}\\
\end{array}\right)
=\left(\begin{array}{cccc}
~\alpha_{\bf k}u_{{\bf k}} & -\beta_{\bf k}u_{{\bf k}} & ~\bar{\sigma}\beta_{\bf k}v^*_{{\bf k}} & -\sigma\alpha_{\bf k}v_{{\bf k}} \\
~\beta_{\bf k}u_{{\bf k}}  & ~\alpha_{\bf k}u_{{\bf k}} & -\bar{\sigma}\alpha_{\bf k}v^*_{{\bf k}} & -\sigma\beta_{\bf k}v_{{\bf k}} \\
-\sigma\beta_{\bf k}v^*_{{\bf k}}  & ~\bar{\sigma}\alpha_{\bf k}v_{{\bf k}} & ~\alpha_{\bf k}u_{{\bf k}} & -\beta_{\bf k}u_{{\bf k}} \\
~\sigma\alpha_{\bf k}v^*_{{\bf k}}  & ~\bar{\sigma}\beta_{\bf k}v_{{\bf k}} & ~\beta_{\bf k}u_{{\bf k}} & ~\alpha_{\bf k}u_{{\bf k}} \\
\end{array}
\right)
\left(\begin{array}{c}
d_{1{\bf k},\sigma}\\
d_{2{\bf k},\bar{\sigma}}\\
d_{1{\bf k}+{\bf Q},\sigma}\\
d_{2{\bf k}+{\bf Q},\bar{\sigma}}\\
\end{array}\right).
\label{Unitary}
\end{eqnarray}
\end{widetext}

An immediate ansatz emerges from the above transformation$-$guided by the same symmetry of the Hamiltonian in Eq.~\ref{Hint}$-$that the system is time-reversal invariant under the representation of this symmetry $\mathcal{T}=\mathbb{I}\otimes i\tau^y\mathcal{K}$, where $\mathcal{K}$ is the complex conjugation. Under time reversal, the Hamiltonian transforms as $H_{\bf k}=\mathcal{T}H_{-{\bf k}}\mathcal{T}^{-1}=\mathbb{I}\otimes\tau^y\left( H^*_{-{\bf k}}\right)\mathbb{I}\otimes \tau^y$. This important symmetry consideration renders a {\it zero} magnetic moment, as depicted schematically in Fig.~\ref{fig2}*b), which is confirmed by a trivial analytical computation of the spin operator: $S=\frac{1}{N}\sum_{i{\bf k}}\langle c^{\dag}_{i{\bf k}+{\bf Q},m}\sigma^l_{mn}c_{i{\bf k},n}\rangle=0$ for all three components of Pauli matrices $\sigma^l$ for $l=x,y,z$ for spin; see Appendix~\ref{Sec:A2}. Here indices $i=1,2$ are for two bands and $m$ and $n$ are components of Pauli matrices, and $N$ is the total number of states. This is due to the fact that magnetic moment in different orbitals cancel each other owing to SOC.

On the other hand, the SODW order parameter has finite expectation value as
\begin{eqnarray}
\sigma \Delta_0&=&\frac{V}{N}\sum_{ij{\bf k}}\left\langle  c^{\dag}_{i{\bf k}+{\bf Q},m}i\tau^y_{ij}\sigma^x_{mn}c_{j{\bf k},n}\right\rangle\nonumber\\
&=&\frac{2V}{N}\sum_{\bf k}^{\prime} u_{\bf k}{\rm Re}\left[v_{\bf k}\right]\left[f(\tilde{E}^+_{\bf k})-f(\tilde{E}^-_{\bf k})\right],
\label{SODW}
\end{eqnarray}
where the fermion occupation number is defined as $f(\tilde{E}^+_{\bf k})=\langle d^{\dag}_{1{\bf k},\sigma}d_{1{\bf k},\sigma}\rangle$ and so on, and $\tilde{E}^{\pm}_{\bf k}$ are the degenerate eigenstates of the interaction Hamiltonian in Eq.~(\ref{Hint}). The notation `prime' over the momentum summation indicates that the summation is restricted within the reduced Brillouin zone. The complex gap parameter that enters in the Hamiltonian is $\Delta=\Delta_0(\sigma_x+\sigma_y)$.  For the experimental value of gap amplitude $\Delta_0\sim 10$~meV, we estimate the critical inter-orbital interaction strength to be $V\sim0.6$~eV which is a reasonable number for 5$f$ electrons estimated earlier.\cite{Dassf}

The free-energy of the HO state is deduced from $F=-k_BT{\rm ln}({\rm Tr} e^{-H/k_BT})+N\mu$, where $k_B$ is Boltzmann constant, $N$ is number of filled states, and $\mu$ is the chemical potential. In the diagonal basis with a mean-field order, the free energy translates into $F=-k_BT\sum^{\prime}_{\bf k} {\rm ln}[\sum_{\nu=\pm}e^{-\tilde{E}_{\bf k}^{\nu}/k_BT}]+N(\mu+|\Delta|^2/V)$. And the corresponding entropy release at the HO transition is evaluated to be $\Delta S=(\partial F/\partial T)_{T_h=17.5~K}\sim0.3k_B{\rm ln}2$, which is close to its experimental estimate from the specific heat jump.\cite{Cv_Palstra}

\begin{figure*}[top]
\rotatebox[origin=c]{0}{\includegraphics[width=1.99\columnwidth]{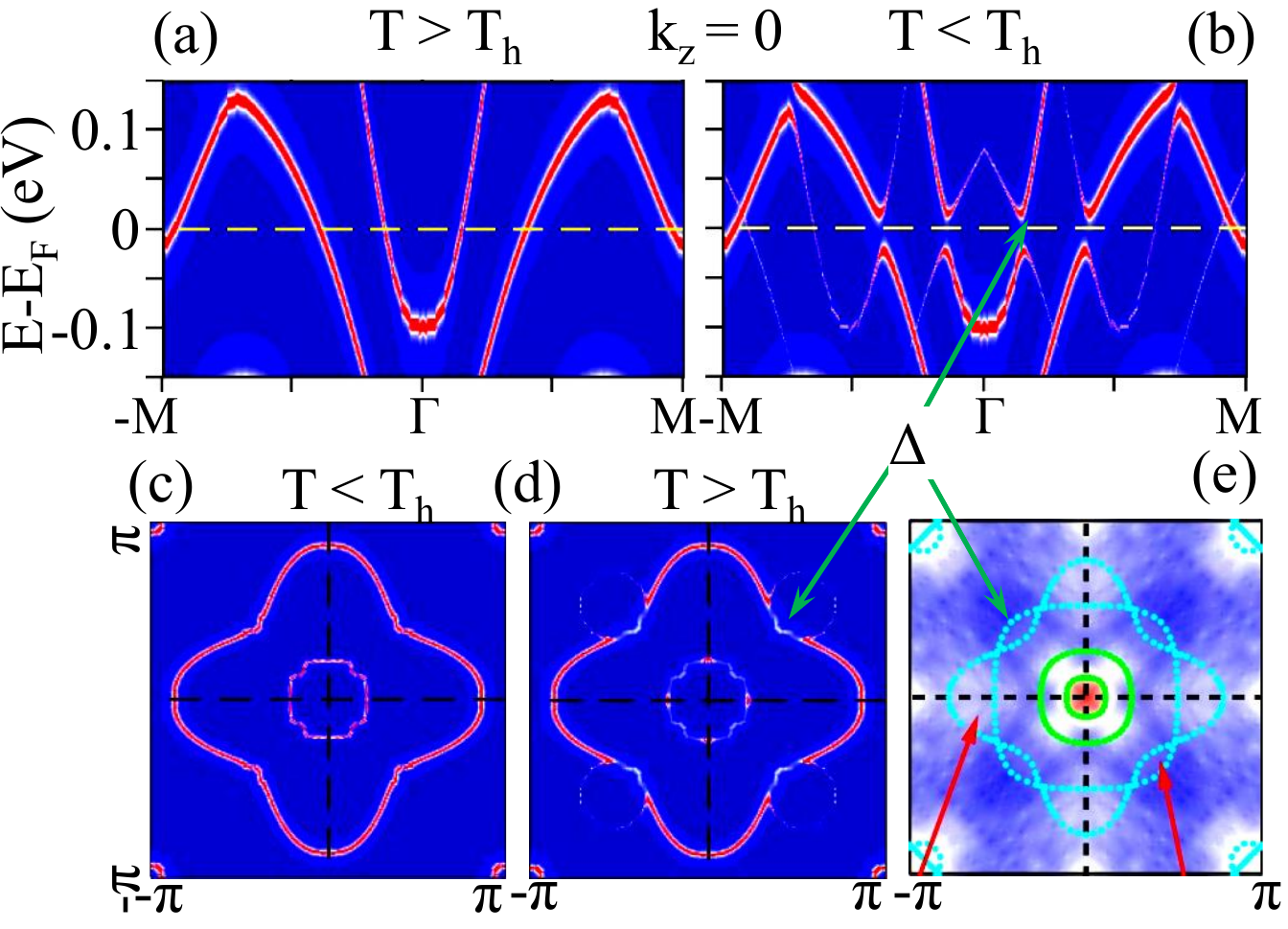}}
\caption{{Images of the  electronic fingerprints of the SODW gap, and the `surprising' collapse of conduction band at the HO transition.} (a) Spectral weight maps of the bare band dispersion along the (110) direction above the HO transition, deduced from DFT calculation with a constant broadening of 0.5~meV. (b) Corresponding quasiparticle dispersion in the HO state within the SODW scenario at the modulation wavevector of ${\bf Q}$. Clear semiconductor-like gap at the Fermi level is observed for the two bands centering $\Gamma$-point (indicated by arrow).  (c)-(d) Computed FS below and above the HO transition on the $k_z=0$ plane, respectively. A good correspondence between theory and experiment can also be marked here. (e) ARPES FS data validates the loss of quasiparticle weight on regions along the (110) direction in BCT crystal structure. The light blue symbols were in the original experimental figure which shows a DFT band structure computed in the ST phase,\cite{DFT} as opposed to the BCT structure of the original lattice. The red arrows are also from the original experimental figure dictating the gapped and ungapped regions.}
\label{fig3}
\end{figure*}

\section{Results}
The state-of-the-art electronic and magnetic fingerprints of the SODW induced HO state are evaluated using DFT-based band structure as the input frame of reference, and writing down the reduced space Hamiltonian with `hot-spot' wavevector ${\bf Q}=(\pi/2,\pi/2,0)$ in the BCT crystal structure. In presenting the results, we focus on the representative electronic properties obtained from ARPES,\cite{Durakiewicz,ARPES,ARPES2} and the spin-excitation spectrum measured by inelastic neutron scattering (INS) experiment.\cite{INS2,INS} Figure~\ref{fig3} gives the single-particle maps along several representative momentum cuts, and the FS topology before and after the HO transition. Gapping of the FS is clearly visible for the two bands aligned along the (110)-direction at the degenerate Dirac points at the Fermi level, introduced in Fig.~\ref{fig1}(a). This gapping process truncates the paramagnetic FS into small pockets aligned along the bond-direction. We require paying particular attention to the location of the FS pockets in this system, since the original crystal structure is BCT (not simple tetragonal which is often used for the simplification of computation). In this unit cell notation, the location of gapping and FS pockets governed in the SODW state are in direct agreement with the ARPES spectral function map on the FS, as shown in Fig.~\ref{fig3}(e). We find that the locii of the gapped states move away from the Fermi level as we increase $k_z$ value. We note that a Shubunikov de-Hass (SdH) measurement finds that the quantum oscillation frequency does not change by any significant amount in going from HO phase at ambient pressure to the large moment antiferromagnetic (LMAF) phase at high pressure,\cite{Hassinger} indicating that the HO phase is intertwined with the LMAF phase at high pressure and at finite magnetic field.

\begin{figure*}[top]
\rotatebox[origin=c]{0}{\includegraphics[width=1.9\columnwidth]{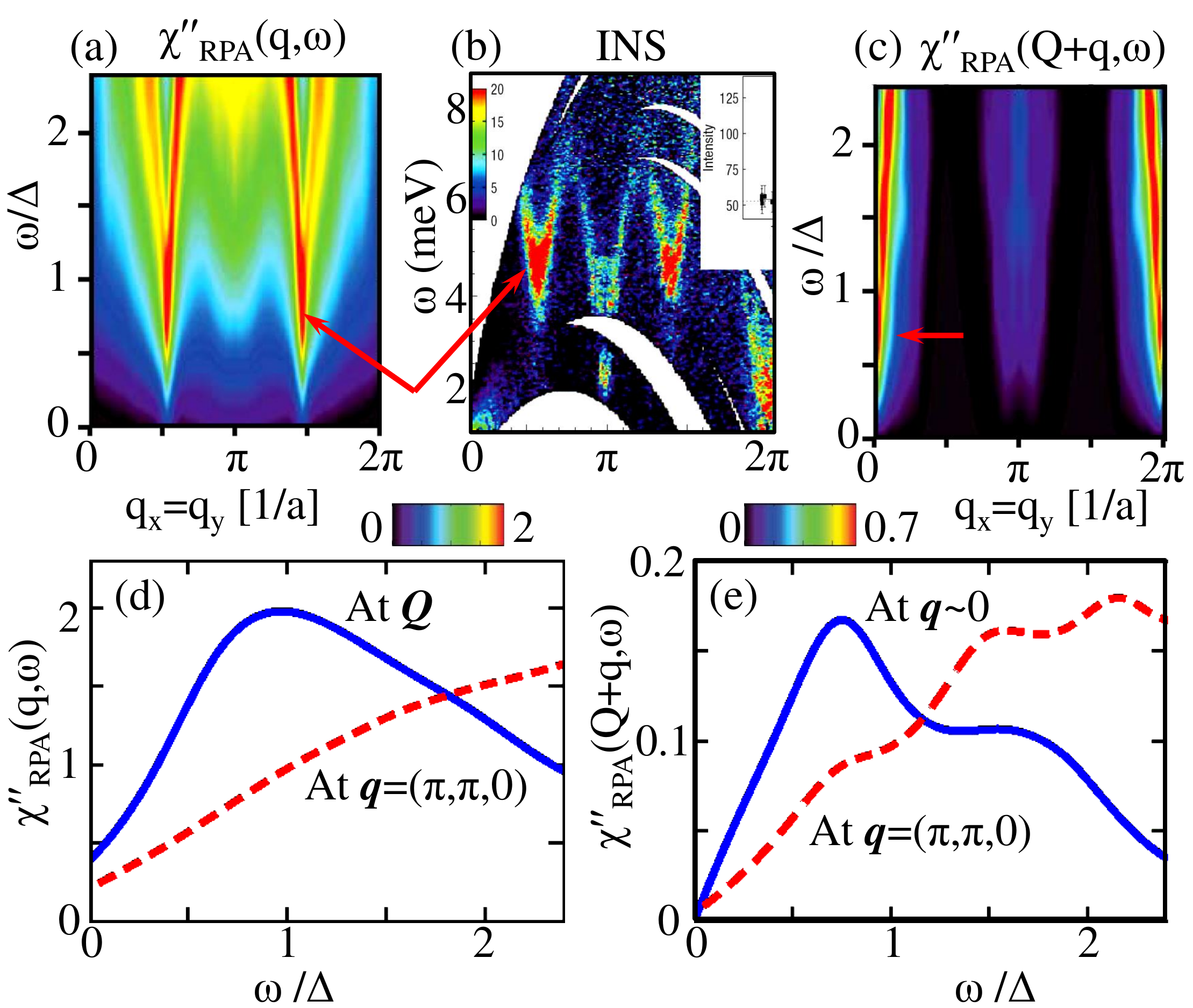}}
\caption{{Computed spin-excitation spectrum within RPA and comparison with available INS data.}
Imaginary part of the spin-excitation spectrum plotted along the ${\bf q}$=(110) direction as a function of excitation energy in the HO state. (a) For the inter-band (inter-orbital) components, we observe the development of a collective mode inside the HO gap with an upward linear-in-energy dispersion starting from ${\bf Q}\sim(0.55\pi,0.55\pi,0)$. Although a good correspondence between theory and experiment is visible when compared with the INS data\cite{INS} in (b), a lattice transformation between BCT to ST is required to align the (100)-direction in the experimental figure to our plot along the (110) direction. The gapped Goldstone mode or roton-mode is well reproduced via the SODW order parameter at this incommensurate wavevector, which deviates slightly from the embedded HO wavevector of ${\bf Q}$ due to the band structure effect. (c) Umklapp spin susceptibility exhibits a second collective mode, localized at ${\bf q}\sim0$ with intensity which is an order of magnitude lower than that for the intra-band one. This mode can be probed via ESR or NQR measurements. (d) $\chi^{\prime\prime}$ is plotted at ${\bf Q}$, and at commensurate ${\bf q}=(\pi,\pi,0)$ (dashed line), taken from (a). (e), Same as (d) but for umklapp contributions at ${\bf q}\sim 0$ and ${\bf q}=(\pi,\pi,0)$, taken from (c).}
\label{fig4}
\end{figure*}

\subsection{Spin-1 collective modes} Our final results of the spin-excitation spectrum have two-fold objectives. They give a realistic fit to the existing data of INS both in dispersion and energy values, and also make a prediction of a second ${\bf q}\sim 0$ mode in the SODW which can be probed by ESR or NQR without any applied static magnetic field. INS measures the imaginary part of the spin susceptibility$-$modulo materials specific form factor$-$which is theoretically computed as a convolution of the single-particle Green's function in the HO state $\chi^{\sigma\bar{\sigma}}_{mn}({\bf q},\omega)=(N\beta)^{-1} \sum_{{\bf k},p}G_m({\bf k},\sigma,i\Omega_p)G_n({\bf k}+{\bf q},\bar{\sigma},i\Omega_p+\omega)$, where $i\Omega_p$ is the fermionic Matsubara frequency and $m,~n$ are band indices, and $G_m({\bf k},\sigma,i\Omega_p)$ is the Green's function in the $m^{th}$ band and so on, and $\beta=1/k_BT$. The corresponding numerical results are obtained by taking analytical continuation to the real frequency axis and integrating over the complex plane as $i\Omega_n\rightarrow\Omega+i\delta$ with $\delta$ being infinitesimal. We include the many-body interaction within the random-phase approximation (RPA), see Appendix~\ref{Sec:A1}. RPA calculation is justified in this heavy-fermion systems at low-$T$ where itinerant electronic structure is well-established. The imaginary part of the RPA susceptibility $\chi_{\rm RPA}$ is presented in Figs.~\ref{fig4}(a) and (c) and compared with corresponding INS spectrum\cite{INS} in Fig.~\ref{fig4}(b). It should be noted that the mechanism and structure of the interacting susceptibility is mainly tied to the details of the electronic structure and HO `hot-spot' wavevector embedded within this bare susceptibility; however, it becomes a collective excitation when RPA correction is included. For a two-band model,  the RPA interaction Hamiltonian (see Appendix~\ref{Sec:A1} for the full expression for any number of orbitals) is
\begin{eqnarray}\label{RPAHam1}
H_{int} &=& \sum_{{\bf k},{\bf k}^{\prime}}\left[\sum_{i} U c^{\dag}_{i{\bf k}\uparrow}c_{i{\bf k}\uparrow}c^{\dag}_{i{\bf k}^{\prime}\downarrow}c_{i{\bf k}^{\prime}\downarrow}+ V c^{\dag}_{1{\bf k}\uparrow}c_{1{\bf k}\uparrow}c^{\dag}_{2{\bf k}^{\prime}\downarrow}c_{2{\bf k}^{\prime}\downarrow}\right].\nonumber\\
\end{eqnarray}
Different bandwidths of different orbitals amount to different critical values of $U$, and $V$, however, to justify that the final results are parameter free, we take the lowest critical value of $U$=1~eV for all low-energy orbitals, and $V$=0.6~eV which shifts the energy scale of all the excitation mode presented below slightly to a lower value.

Figure~\ref{fig4}(a) presents the computed inter-band spin excitation spectrum along the zone diagonal direction, and compares with the corresponding experimental data\cite{INS} obtained along the zone boundary direction given in Fig.~\ref{fig4}(b). For the inter-band transition in the particle-hole continuum, we clearly mark a prominent ${\bf S}=1$ collective mode with a dispersion that resembles a gapped Goldstone mode or a roton-like spectrum. For the case of the discrete symmetry breaking due to the complex order parameter $\Delta^*$, such gapping of the Goldstone mode is expected. The mode is localized around ${\bf Q}\sim(0.55,0.55,0)\pi$, slightly shifted from the original HO wavevector ${\bf Q}$ due to the band structure effect. However, the shift of the mode below $\omega<\Delta$ is a many-body correction as discussed above. To match the mode energy of the experimental data about $\omega$=5~meV, a gap amplitude of $\Delta_0\sim 6$~meV is invoked in our calculation which is close to the spectroscopic value of the HO gap.\cite{STM,STM2} Although a static neutron signal is reported in Ref.~\cite{zeromoment} at ${\bf q}=(\pi,\pi,0)$, but in the inelastic spectrum, both experiment\cite{INS} and our theory find a weaker and featureless intensity at this wavevector, as compared in Fig.~\ref{fig4}(d). We note that the INS spectral weight loss at the incommensurate wavevector, as opposed to the commensurate one, can fully account for the entropy loss at the
HO transition.

By construction, the SODW is associated with an interaction induced spin-orbit entangled electronic structure, rendering conceptual  similarity with the spin-orbit order in the particle-particle channel proposed by Leggett\cite{Leggett} for the liquid $^3$He superconducting phase, dynamic generation of SOC\cite{CWu}  or the single-particle quantum spin-Hall state\cite{SCZSHE} which do not break time-reversal symmetry. In what follows, the umklapp scattering process between different spin-orbital states brings out two Zeeman-like spin-split states at the same momentum {\bf k}, on both sides of the HO gap with coherence factors $u_{\bf k}/v_{\bf k}$, without any external magnetic field, as discussed in Fig.~\ref{fig2}(a). Therefore, a second spin-flip collective mode is expected in the off-diagonal susceptibility $\chi^{\sigma\bar{\sigma}}_{12}({\bf q}+{\bf Q},\omega) \sim \sum_{\bf k}\delta(\omega-\xi_{1{\bf k}}+\xi_{2{\bf k}+{\bf Q}+{\bf q}})$ to localize inside the SODW gap $\omega\le|\Delta|$ around ${\bf q}\sim 0$. Indeed, our computation confirms the existence of this collective mode at energy $\omega/\Delta\sim0.75$, as shown in Figs.~\ref{fig4}(c) and \ref{fig4}(e). Since the intensity of this mode is about an order of magnitude lower than that of the ${\bf Q}$-mode presented in Fig.~\ref{fig4}(a), it will be difficult to simultaneously detect them both in the INS measurements. However, ESR or NQR, having the capability of detecting lineshifts of resonance without the application of a static magnetic field, will be able to measure our proposed ${\bf q}\sim0$ mode.

In addition, we also reproduce a large anisotropy in the spin-susceptibility with the anisotropy arising below $T\sim 120$~K. This implies that the susceptibility anisotropy is related to the bandstructure of this compound, and not triggered by the HO phase. Our result (not shown) is similar to the one obtained earlier in Ref.~\onlinecite{TheoryIkeda} using similar DFT bandstructure, and agrees well with experiment.\cite{Cv_Palstra}

\section{Conclusions}
Broken and invariant symmetries drive quantum phase transitions and the stability of a phase, respectively, and thus the uncharted physical phenomena emerging from their interplay can inherit even more interesting merits. Spin-orbit entanglement has gained recent attention along this line which has been shown to foster the quantum spin-Hall state,\cite{SCZSHE} and other topologically protected phases\cite{CKane,SCZhang} due to the time-reversal invariance. It is also shown that the spin-orbit coupling is directly or indirectly responsible for causing Landau-type novel quantum phase transitions via, for example, translational symmetry breaking where time-reversal symmetry is not necessarily broken in two-dimensional electron gas,\cite{DasPRL} and iridates\cite{Iridates}. SODW is a leading example of this kind that we formulate and characterize here.\cite{DasSR} The direct consequence of this order is that the magnetic field is a perturbation to this order (for symmetry reason), in addition to the temperature. Of course, since the SODW renders FS gapping, and thereby larger resistivity with respect to its paramagnetic value, the application of magnetic field will lead to a reduction in resistivity (when other parameters are kept constant) which is often refereed as the negative magnetoresistance effect. Finally, the shared coherence weight of the quasiparticle weight between the spin-orbit entangled state leads to a  collective excitation mode, localized in energy $\omega<\Delta$ at the zero momentum transfer. This is shown here to exist in the URu$_2$Si$_2$ band structure below the HO transition, and can be measured in future ESR or NQR or polarized INS measurements. Taken together, a generalized framework for the theory of SODW and its relevant electronic and magnetic fingerprints are deduced here and shown to reconcile a number of experimental signatures of the HO state in URu$_2$Si$_2$ which were taken earlier to be contradictory according to conventional theories. The essential ingredient for the realization of a SODW in any other systems is that the single-electron wavefunction inherits a spin-orbit coupling stronger than its effective inter-orbital Coulomb interaction. Finally, we note that the possible competition and/ or coexistence of the HO phase with large moment antiferromagnetic phase at high-pressure will be studied in a future work.\cite{DasIridates}

\begin{acknowledgments}
The author acknowledges valuable discussion with A. Leggett, J. Mydosh, P. Coleman, M. J. Graf, P. Woelfle, and J.-H. She. The author expresses gratitude to A. Leggett for suggesting the study of ESR for the SODW order parameter, and to Jian-Xin Zhu for sharing some of the first-principles band structure files for Wien2K calculation. The work is supported by the U.S. DOE through the Office of Science (BES) and the LDRD Program and facilitated by NERSC computing allocation.
\end{acknowledgments}

\appendix
\section{Interaction vertex of the SODW order and RPA correction}\label{Sec:A1}

In this section we identify interaction terms that contribute to the SODW vertex. In the multiorbital setup, the full interacting Hamiltonian includes intra- and interorbital Coulomb interactions, $U$ and $V$, Hund's coupling $J$, and pair-exchange term $J^{\prime}$:
\begin{eqnarray}
H_{int}&=&\sum_{{\bf k}_1-{\bf k}_4}\left[\sum_i U c^{\dag}_{i{\bf k}_1,\uparrow}c_{i{\bf k}_2,\uparrow}c^{\dag}_{i{\bf k}_3,\downarrow}c_{i{\bf k}_4,\downarrow}\right.\nonumber\\
&&+\sum_{i<j,\sigma}\left(Vc^{\dag}_{i{\bf k}_1,\sigma}c_{i{\bf k}_2,\sigma}c^{\dag}_{j{\bf k}_3,\bar{\sigma}}c_{j{\bf k}_4,\bar{\sigma}}\right.\nonumber\\
&&~~~+\left.(V-J)c^{\dag}_{i{\bf k}_1,\sigma}c_{i{\bf k}_2,\sigma}c^{\dag}_{j{\bf k}_3,\sigma}c_{j{\bf k}_4,\sigma}\right)\nonumber\\
&&+\sum_{i<j,\sigma}\left(Jc^{\dag}_{i{\bf k}_1,\sigma}c^{\dag}_{j{\bf k}_3,\bar{\sigma}}c_{i{\bf k}_2,\bar{\sigma}}c_{j{\bf k}_4,\sigma}\right.\nonumber\\
&&~~~+\left.\left. J^{\prime} c^{\dag}_{i{\bf k}_1,\sigma}c^{\dag}_{i{\bf k}_3,\bar{\sigma}}c_{j{\bf k}_2,\bar{\sigma}}c_{j{\bf k}_4,\sigma} + h.c.\right)\right].
\label{intH}
\end{eqnarray}
where ${\bf k}_1+{\bf k}_3={\bf k}_2+{\bf k}_4$. The linearized gap equation for the SODW order parameters is $1=-T_h\Gamma({\bf Q},\omega)$ where $T_h$ is the SODW transition temperature. Considering all possible contractions of the interaction terms written in Eq.~(\ref{intH}), we get the SODW vertex as\cite{DasIridates}
\begin{eqnarray}
\Gamma({\bf Q},\omega) &=& V{\tilde \chi}_{12}^{12}({\bf Q},\omega) + J^{\prime}{\tilde \chi}_{12}^{21}({\bf Q},\omega),
\label{sceqn}
\end{eqnarray}
 where the corresponding non-ineracting susceptibilities are defined as
\begin{eqnarray}
{\tilde \chi}_{il}^{jk}({\bm q},\omega) &=& -\frac{1}{N\beta}\sum_{{\bm k},\Omega_p}G_{ij}({\bm k},i\Omega_p)G_{kl}({\bm k}+{\bm q},i\Omega_p+\omega),\nonumber\\
\label{chi}
\end{eqnarray}
with $i,j$ and $k,l$ being orbital indices for the initial and final states in the scattering process. $G_{ij}$ is the single-particle Green's function in the orbital basis (while that presented in the main text is in the band basis). ${\tilde \chi}_{il}^{jk}$ is a matrix of dimension $n^2\times n^2$, where $n$ is the number of orbitals.\cite{DasIridates}  In this definition, $\chi_{11}^{11}$, $\chi_{12}^{12}$ are the intra- and inter-orbital components of the susceptibility, and $\chi_{11}^{22}$, $\chi_{12}^{21}$ are Hund's and pair scattering susceptibilities, respectively, and so on.\cite{DasIridates} In general, the pair scattering term is small compared to the other interactions, and thus for many practical purposes it can be neglected.  However, we do not find that the Hund's coupling $J$ contributes to the SODW order which we envisaged in Ref.~\cite{DasSR}. Hund's coupling may contribute to the inter-orbital spin density wave as proposed by Riseborough {\it et al.}\cite{Riseborough} For $J=J^{\prime}=0$, this interacting Hamiltonian reduces to Eq.~(\ref{RPAHam1}).

Finally, in the SODW state, each component of non-interacting susceptibility again becomes a 2$\times$2 matrix, made of direct transitions (in the diagonal terms) and umklapp scattering terms placed in off-diagonals. The direct term means scattering between two main bands and between two shadow bands while the umklapp term arises from the scattering between main and shadow bands. The direct term gives a gapped spectrum at the `hot-spot' vector, while the umklapp term is responsible for the ${\bf q}\sim 0$ mode as discussed  in the main text.

The interacting susceptibility is calculated within RPA framework in which the RPA susceptibility matrix is ${\tilde \chi}_{\rm RPA}={\tilde \chi}^0/(\mathbb{I}-{\tilde \Gamma}_{\rm RPA}{\tilde \chi})$, where $\mathbb{I}$ is the identity matrix and ${\tilde \Gamma}_{\rm RPA}$ is the RPA vertex of same dimension as of ${\tilde \chi}$. The component of the RPA vertex matrix should be same as that of the non-interacting susceptibility ${\tilde \chi}$, which means, intra-orbital term ${\tilde \Gamma}_{\rm RPA,11}^{22}=U$, inter-orbital vertex ${\tilde \Gamma}_{\rm RPA,12}^{12}=V$, Hund's coupling ${\tilde \Gamma}_{\rm RPA,11}^{22}=J$, and pair-scattering ${\tilde \Gamma}_{\rm RPA,12}^{12}=J^{\prime}$, and so on. Different bandwidths of different orbitals amount to different critical values of $U$, and $V$, however, to justify that the final results are parameter free, we take the lowest critical value of $U$=1~eV for both orbitals, and $V$=0.6~eV which shifts the energy scale of all the excitation mode slightly to a lower energy. As mentioned before, the small contributions of Hund's coupling and pair-scattering terms are neglected here.

\section{Zero magnetic moment}\label{Sec:A2}
The time-reversal symmetry of the SODW Hamiltonian enforces the total magnetic moment to vanish at each spatial point. This can be verified by a trivial analytical computation of the spin-operator:
\begin{eqnarray}
S&=&\frac{1}{N}\sum_{i{\bf k}}\langle c^{\dag}_{i{\bf k}+{\bf Q},m}\sigma^l_{mn}c_{i{\bf k},n}\rangle\nonumber\\
&=&\frac{1}{N}\sum_{i{\bf k}}\alpha_{\bf k}\beta_{\bf k}\left[\sigma\left(-u_{\bf k}v^*_{\bf k}+u_{\bf k}v^*_{\bf k}\right)\langle d^{\dag}_{1{\bf k},\sigma}d_{1{\bf k},\sigma}\rangle \right.\nonumber\\
&&\hspace{0.5cm}\left.+ \bar{\sigma}\left(-u_{\bf k}v_{\bf k}+u_{\bf k}v_{\bf k}\right)\langle d^{\dag}_{2{\bf k},\bar{\sigma}}d_{2{\bf k},\bar{\sigma}}\rangle + \left({\bf k}\rightarrow {\bf k}+{\bf Q}\right)\right] \nonumber\\
&=& 0.
\label{magmoment}
\end{eqnarray}
Here $l=x,y,z$ for three Pauli matrices and $m,n$ are their components. It is interesting to see that for both cases of $\bar{\sigma}=\pm\sigma$, the magnetic moment vanishes, which means there is no magnetic moment associated with the SODW order parameter in any spatial direction. In this spirit, the SODW can be thought of as a `modulated' quantum spin Hall state in which in each sublattice a chiral spin-current is present which becomes flipped in the other sublattices, but there is no net moment induced the system.


\begin{thebibliography}{10}
\bibitem{Cv_Palstra}
T. T. M. Palstra,  A. A. Menovsky, J. van den Berg, A. J. Dirkmaat, P. H. Kes, G. J. Nieuwenhuys, and J. A. Mydosh,
{Phys. Rev. Lett.} {\bf 55}, 2727 (1985).
%
\bibitem{Mydosh_review}
J. A. Mydosh, and P. M. Oppeneer,
{Rev. Mod. Phys.} {\bf 83}, 1301 (2011).
%
\bibitem{INS2}
C.  Broholm, H. Lin, P. T. Matthews, T. E. Mason, W. J. L. Buyers, M. F. Collins, A. A. Menovsky, J. A. Mydosh, and J. K. Kjems
{Phys. Rev. B} {\bf 43}, 12809 (1991).

\bibitem{INS}
C. R. Wiebe, J. A. Janik, G. J. MacDougall, G. M. Luke, J. D. Garrett, H. D. Zhou, Y.-J. Jo, L. Balicas, Y. Qiu, J. R. D. Copley, Z. Yamani, and W. J. L. Buyers,
{Nature Phys.} {\bf 3}, 96 (2007).
%
\bibitem{NMR}
S. Kambe, Y. Tokunaga, H. Sakai, T. D. Matsuda, Y. Haga, Z. Fisk, and R. E. Walsted,
{Phys. Rev. Lett.} {\bf 110}, 246406 (2013).
%
\bibitem{Matsuda}
R. Okazaki, T. Shibauchi, H. J. Shi, Y. Haga, T. D. Matsuda, E. Yamamoto, Y. Onuki, H. Ikeda, and Y. Matsuda
{Science} {\bf 331}, 439 (2011).
%
\bibitem{susceptibility}
A. P. Ramirez, P. Coleman, P. Chandra, E. Brück, A. A. Menovsky, Z. Fisk, and E. Bucher,
{Phys. Rev. Lett.} {\bf 68}, 2680 (1992).
%
\bibitem{zeromoment}
P. Das, R E Baumbach, K Huang, M B Maple, Y Zhao, J S Helton, J W Lynn, E D Bauer and M Janoschek,
{New J. Phys.} {\bf 15}, 053031 (2013).
%
\bibitem{DFT}
S. Elgazzar, J. Rusz,  M. Amft, P. M. Oppeneer, and J. A.  Mydosh,
{Nat. Mat.} {\bf 8}, 337-341 (2009).

\bibitem{TheoryKotliar}
K. Haule, and G.  Kotliar,
{Nature Phys.} {\bf 5}, 796-799 (2009).
%
\bibitem{TheoryPepin}
C. P\'epin, M. R. Norman, S. Burdin,  and A. Ferraz,
{Phys. Rev. Lett.} {\bf 106}, 106601-106604 (2011).
%
\bibitem{TheorySasha}Y. Dubi, and A. V. Balatsky,
{Phys. Rev. Lett.} {\bf 106}, 086401-086404 (2011).
%
\bibitem{TheoryFujimoto} S. Fujimoto,
{Phys. Rev. Lett.} {\bf 106}, 196407-196410 (2011).
%
\bibitem{TheoryHYKee}
J. G. Rau, and H.-Y. Kee,
{ Phys. Rev. B} {\bf 85}, 245112 (2012).
%
\bibitem{TheoryIkeda} H. Ikeda,  {\it et al.}
{Nat. Phys.} {\bf 8}, 528-533 (2012).
%
\bibitem{Riseborough}
P. S. Riseborough, B. Coqblin, and S. G.  Magalh\"aes,
{Phys. Rev. B} {\bf 85}, 165116 (2012).

\bibitem{TheoryHastatic}
P. Chandra,	 P. Coleman,	 and R. Flint,
{Nature} {\bf 493}, 621-626 (2013).
%
\bibitem{SCZSHE}
B. A. Bernevig, T. L Hughes, ans S.-C. Zhang,
{Science} {\bf 314}, 1757 (2006).
%
\bibitem{CKane}
M. Z. Hasan,  and   C. L.  Kane,
{Rev. Mod. Phys.} {\bf 82}, 3045-3067 (2010).
%
\bibitem{SCZhang}
X.-L. Qi, and S.-C. Zhang,
{Rev. Mod. Phys.} {\bf 83}, 1057-“1110 (2011).
%
\bibitem{Weyl}C. Herring,
{Phys. Rev.} {\bf 52}, 365-373 (1937).

\bibitem{Murakami}S. Murakami,
{New J. Phys.} {\bf 9}, 356 (2007).
%
\bibitem{DasSR}
T. Das,
{Sci. Rep.} {\bf 2}. 596 (2012).
%
\bibitem{DasPRL}
T.  Das,
{Phys. Rev. Lett.} {\bf 109}, 246406 (2012).
%
\bibitem{magfield}
J. Levallois, K. Behnia, J. Flouquet, P. Lejay, and C. Proust,
{Europhys. Lett.} {\bf 85}, 27003 (2009).
%
\bibitem{STM}
A. R. Schmidt, M. H. Hamidian, P. Wahl, F. Meier, A. V. Balatsky, J. D. Garrett, T. J. Williams, G. M. Luke, and J. C. Davis,
{ Nature} {\bf 465}, 570-576 (2010).
%
\bibitem{STM2}
P. Aynajian, E. H. da Silva Neto, C. V. Parker, Y. Huang, A. Pasupathy, J.Mydosh, and A. Yazdani,
{Proc. Nat. Acad. Sci. USA} {\bf 107}, 10383 (2010).
%
\bibitem{Wien}
P. Blaha, K. Schwarz, G. K. H. Madsen, D. Kvasnicka, and J. Luitz, {\it WIEN2k, An Augmented-Plane-Wave  +  Local Orbitals Program for Calculating Crystal Properties} (Austria: Karlheinz Schwarz, Techn Wien) (2001).
%
\bibitem{GGA}
J. P. Perdew, K. Burke, and  M. Ernzerhof,
{Phys. Rev. Lett.} {\bf 77}, 3865 (1996).
%
\bibitem{Durakiewicz}
J.-Q. Meng, P. M. Oppeneer, J. A. Mydosh, P. S. Riseborough, K. Gofryk, J. J. Joyce, E. D. Bauer, Y. Li, and T. Durakiewicz,
{Preprint available at http://arxiv.org/abs/1302.4508} (2013).
%
\bibitem{Dassf}
T. Das, J.-X. Zhu, and M. J. Graf,
{Phys. Rev. Lett.} {\bf 108}, 017001 (2012).
%
\bibitem{ARPES}
A. F. Santander-Syro,
{Nature Phys.} {\bf 5}, 637 (2009).
%
\bibitem{ARPES2}
S. Chatterjee, J. Trinckauf, T. Hanke, D. E. Shai, J. W. Harter, T. J. Williams, G. M. Luke, K. M. Shen, and J. Geck,
{Phys. Rev. Lett.} {\bf 110}, 186401 (2013).
%
\bibitem{Hassinger}
E. Hassinger, G. Knebel, T. D. Matsuda, D. Aoki, V. Taufour, and J. Flouquet, Phys. Rev. Lett. {\bf 105}, 216409 (2010).

\bibitem{Leggett}
A. J. Leggett,
{J. Phys. C: Solid State Phys.} {\bf 6}, 3187-3204 (1973).

\bibitem{CWu}C. Wu, and S.-C. Zhang,
Phys. Rev. Lett. {\bf 93}, 36403(2004).

\bibitem{Iridates}
B. J.  Kim, H. Jin, S. J. Moon, J.-Y. Kim, B.-G. Park, C. S. Leem, J. Yu, T. W. Noh, C. Kim, S.-J. Oh, J.-H. Park, V. Durairaj, G. Cao, and E. Rotenberg,
{Phys. Rev. Lett.} {\bf 101}, 076402 (2008).

\bibitem{DasIridates} T. Das, {\it et al.} manuscript under preparation.

\end{thebibliography}
\end{document}